\documentclass[conference]{IEEEtran}
\IEEEoverridecommandlockouts
\usepackage{cite}
\usepackage{amsmath,amssymb,amsfonts}
\usepackage{algorithm}
\usepackage{algorithmic}
\usepackage{graphicx}
\usepackage{textcomp}
\usepackage{xcolor}
\usepackage{caption}
\usepackage{subfigure}
\usepackage{diagbox}
\usepackage{enumerate}
\usepackage{booktabs}
\usepackage{soul}
\def\BibTeX{{\rm B\kern-.05em{\sc i\kern-.025em b}\kern-.08em
    T\kern-.1667em\lower.7ex\hbox{E}\kern-.125emX}}
\begin{document}

\title{SpaceMeta: Global-Scale Massive Multi-User Virtual Interaction over LEO Satellite Constellations}

\author{
 \IEEEcompsocitemizethanks{This work is supported by the National Natural Science Foundation of China (62302292) and the Fundamental Research Funds for the Central Universities. Corresponding author: Yifei Zhu}

\IEEEauthorblockN{Jiahe Huang,
Yifei Zhu}
\IEEEauthorblockA{UM-SJTU Joint Institute,
Shanghai Jiao Tong University\\
Cooperative Medianet Innovation Center(CMIC), Shanghai Jiao Tong University\\
Email: sevenkishuang@sjtu.edu.cn, yifei.zhu@sjtu.edu.cn}}

\maketitle

\begin{abstract}
Low latency and high synchronization among users are critical for emerging multi-user virtual interaction applications. However, the existing ground-based cloud solutions are naturally limited by the complex ground topology and fiber speeds, making it difficult to pace with the requirement of multi-user virtual interaction. The growth of low earth orbit (LEO) satellite constellations becomes a promising alternative to ground solutions. To fully exploit the potential of the LEO satellite, in this paper, we study the satellite server selection problem for global-scale multi-user interaction applications over LEO constellations. We propose an effective server selection framework, called SpaceMeta, that jointly selects the ingress satellite servers and relay servers on the communication path to minimize latency and latency discrepancy among users. Extensive experiments using real-world Starlink topology demonstrate that SpaceMeta reduces the latency by 6.72\% and the interquartile range (IQR) of user latency by 39.50\% compared with state-of-the-art methods.

\end{abstract}


\section{Introduction}

In recent years, multi-user virtual interaction finds widespread applications in online games and social networking applications, where users constantly interact with thousands of other users via virtual avatars, video streaming, and audio. 

For example, platforms like Facebook Spaces and VRChat enable users to connect with friends, explore virtual environments, and partake in diverse activities using personalized avatars. It is reported that the 24-hour average player number on Steam-based VRChat already surpassed 20,000 in June 2023 \cite{VRChat}. In general, over 15.49 million consumer VR headsets are sold in 2022 \cite{Statista}.

The increasing popularity of multi-user virtual interaction demands a seamless user experience for a massive number of users. To provide an immersive experience to these users, low latency, high synchronization among users, as well as high scalability of the system are crucial. Low latency is critical to enabling real-time interactions; High synchronization ensures consistent and coordinated events, actions, and feedback for all users. Accommodating as many users as possible improves the social value of the applications and consequently the quality of experience for every user within the system for the better interaction feature. 

Current architectures of multi-user interaction applications predominantly rely on ground cloud servers. However, in such an architecture, terrestrial communications tend to travel through optical fibers, which are approximately 30\% slower than the speed of light. This essentially limits the data transmission speed of cloud-based interactions. Additionally, the complex topography of the Earth's surface, such as across oceans, often leads to longer transmission paths, further increasing latency.

With the advancement of space techniques and the maturity of low-cost communication satellites, LEO satellites emerge as a promising solution to support global-scale communication. 
LEO satellites offer enhanced communication capabilities, allowing information to travel through space close to the speed of light, with fewer spatial constraints. Thousands of LEO satellites further form a satellite network, also known as constellations, where satellites work together to provide extensive coverage. As of May 2023, Starlink comprises over 4000 LEO satellites of different phases and groups.

These satellites are further connected by inter-satellite links (ISLs). ISLs establish direct connections between satellites, eliminating the need for ground-based infrastructure and reducing reliance on traditional terrestrial communication networks. By leveraging ISLs, satellite constellations like Starlink achieve faster and more efficient data transfer, bolster network resilience, and enhance global coverage. This revolutionizes satellite communications and unlocks new possibilities for global connectivity. With a substantial number of satellites in orbit and wide coverage, LEO satellites can serve as servers and control units in multi-user virtual interaction architectures. The LEO mega-constellation holds significant promise in meeting this need. 

However, moving multi-user virtual interaction applications from the ground to space poses significant challenges. Firstly, the rapid movement of satellites introduces a dynamic topology, requiring adaptive strategies to maintain a stable connection. The continuous access and departure of users during the session add another layer of complexity.  Second, the ingress satellite server and the communication path between communication pairs have to be carefully selected so that the latency can be minimized and the synchronization among all pairs can be maximized. 

In this paper, we study the satellite server selection problem for global-scale multi-user interaction over LEO satellite constellations. We first present an architecture that utilizes satellites as relay servers. We then propose SpaceMeta that intelligently selects ingress server and relay servers jointly minimize the latency and the latency discrepancy among users in a virtual interaction session. Extensive experiments on real-world application scales demonstrate the efficiency of our approach.

In summary, our contributions are: 
\begin{itemize}
    \item We present the first work to examine the possibility of supporting global-scale multi-user virtual interaction over the emerging LEO satellite constellations. 
    \item We formulate the satellite server selection problem as a latency and synchronization minimization problem. 
    \item We propose an effective greedy solution to identify the first hop ingress server in the LEO constellations as well as the transmission path between users. 
\item Extensive experiments on the scale of the Starlink constellation demonstrate that our approach can reduce latency by up to 40.67\% and interquartile range of latency by up to 80.28\% when compared with the state-of-the-art solutions.  
\end{itemize}

The rest of the paper is organized as follows: Section \ref{related work} presents the related work. Section \ref{system model} introduces the system model of SpaceMeta and formulates the problem. Section \ref{method} showcases the relay selection and flow generation algorithms involved. Section \ref{results} includes the test results of our approach, which are also compared with other existing methods. Finally, Section \ref{conclusion} concludes the work.

\section{Related Work}\label{related work}
\subsection{Satellite Networks and Applications}

LEO satellites offer a promising network establishment approach to manage the growing number of edge devices and offer global communication coverage. In \cite{starperf simulator}, StarPerf employs AGI STK \cite{STK} to simulate motion trajectories of Starlink (Phase 1) satellites and assess inter-satellite connectivity based on distance. In \cite{SpaceRTC}, a cloud-satellite combined architecture, incorporating both LEO satellites and ground base stations as potential servers, is proposed to reduce average attendee latency. UAVs (unmanned aerial vehicles) can also act as links between satellites and ground stations. Researchers in \cite{UAV} explore UAVs as relay servers, efficiently gathering and transmitting IoT (Internet of Things) data from numerous IoT devices to LEO satellites. In hybrid satellite-terrestrial modes, Ma et al. in \cite{Reliability} investigate reliability issues in long-distance relaying transmissions, considering end-to-end latency and outage performance with geometrical probability theory.

In addition to latency, numerous studies have also explored additional indicators such as bandwidth, power, and throughput in data transmission. For power allocation in relay satellites to maximize total data rate, Wang et al. in \cite{maximum rate} utilize virtualization, the difference of convex (DC) programming, and an iterative algorithm to achieve optimal solutions. Addressing limited spectrum resources and the demands of numerous users, researchers in \cite{NOMA} establish cooperative transmission of non-orthogonal multiple access-assisted integrated satellite-terrestrial networks with multiple terrestrial relays and consider a partial relay selection scheme. 

Furthermore, researchers in \cite{Throughput} tackle the conflict between latency and throughput in relay selection, formulating antenna scheduling as a stochastic non-convex fractional programming and transforming it into a solvable weight-matching problem through mathematical methods.

\subsection{Multi-user interactive systems}

Numerous studies focus on low-latency multi-media systems, addressing architecture, coding, and transport protocols. For instance, researchers in \cite{J-QoS} introduce a novel approach that utilizes application flow destinations to calculate service delay. They implement forwarding, caching, and innovative coding services, reducing packet recovery costs by combining packets from multiple application streams and strategically sending a limited number of coded packets via more expensive cloud paths. In the same vein, Salsify in \cite{Salsify} optimizes compressed frame length and transmission time based on network capacity estimation, enabling real-time Internet video to promptly respond to network changes, prevent packet drops, and avoid queuing delays. Moreover, VIA in \cite{VIA} is an architecture that utilizes selected ground station-based relays to construct a managed overlay network, effectively reducing latency in Internet telephony. Notably, the three systems discussed are all cloud-based only.

Multi-user virtual interaction is an extensively researched domain. Chen et al. in \cite{metaverse architecture} present fundamental metaverse concepts, including computing, logical, physical, and protocol architectures. The study emphasizes the significance of edge computing and edge server placement in optimizing resource utilization and access delay. Meanwhile, researchers in \cite{metaverses} address issues such as limited user vision in the metaverse by implementing a server partitioning system based on regions, enhancing the object discovery service, and ensuring minimum throughput for communicating objects. For a broader perspective, Solipsis in \cite{MMVE} tackles the world-scale Metaverse using a Raynet-based decentralized architecture protocol, considering geometric distributions in the real world. Moreover, researchers in \cite{edge intelligence-based} optimize communication and computing in the digital twins-enabled metaverse, utilizing edge servers to reduce latency and enhance stability. Additionally, SLAMCast in \cite{SLAMCast} is an innovative MC-based transmission protocol, which effectively lowers bandwidth requirements in multi-user VR scenes, and achieves a stable client-server system with the aid of a novel thread-leveled GPU hash set. 

In multi-user and multi-party interactive live streaming, ensuring synchronization among all attendees holds significant importance. RobinHood in \cite{osdi} addresses this concern by concentrating on reducing tail latency in CDN-based video streaming systems. The work proposes a novel approach involving the dynamic reallocation of cache resources, effectively repurposing existing caches to mitigate the impact of backend latency variability, resulting in low tail latency.

\section{System Model and Problem Formulation}\label{system model}

\begin{figure}[!tbp]
    \centering
    \includegraphics[width=3.0in]{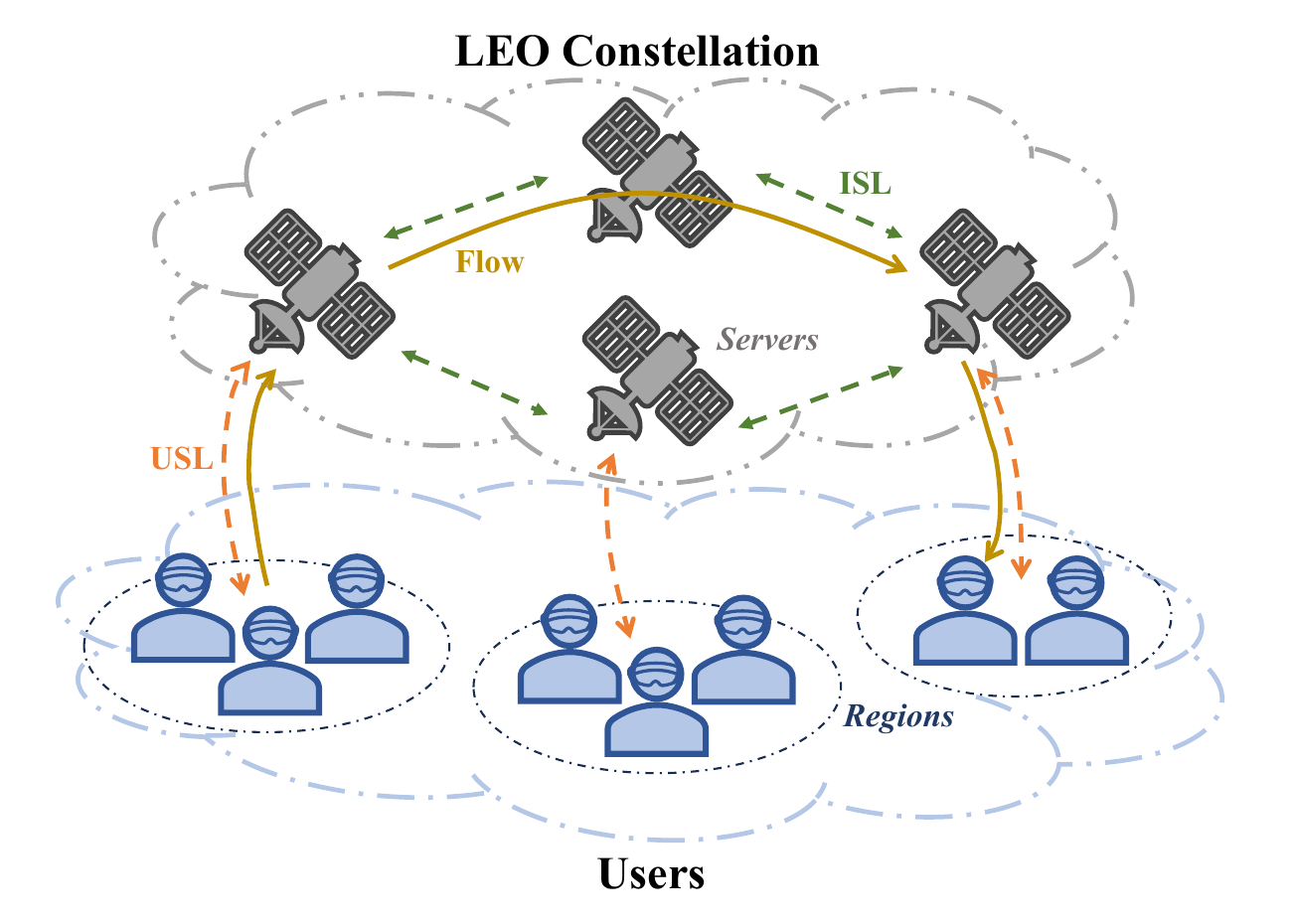}
    \caption{System overview.}
    \label{architecture}
\end{figure}

\subsection{System Overview}
Figure \ref{architecture} depicts the overall architecture of the multi-user interaction application supported by the LEO constellation. The architecture is divided into two primary segments: the space segment and the ground segment. In the space segment, LEO mega-constellations like Starlink are utilized, establishing inter-satellite links (ISLs), and the satellites serve as servers, including ingress servers and others. On the other hand, the ground segment encompasses all users, classified into different regions based on population distribution, with each region assigned a specific ingress server. The two segments are interconnected through user-to-satellite links (USLs), and the flow between users is governed by algorithms detailed in \ref{method}.

\subsection{Connection Model}
 Denote $\mathbb{S}$ as the set of all satellites. The multi-user virtual interaction consists of various sessions, denoted as $\mathbb{E}$. Within a specific session $e$, we use $\mathbb{U}^e$ to represent the set of all users. The system comprises two types of links: inter-satellite links (ISLs), represented as $ISL(m, n)$, where $ m,n\in\mathbb S$, and user-satellite links (USLs), denoted as $USL(m, n)$, where $m\in\mathbb U, n\in\mathbb S$.

The movement of LEO satellites at high speeds impacts the link between two satellites due to their visibility. To address this, we introduce a binary visibility variable denoted as $Vis(m,n)$, where $Vis(m,n)=Vis(n,m)=1$ if and only if node $m$ and $n$ are visible to each other. Typically, visibility is directly related to the distance and is achievable only between adjacent satellites. For USLs, a link is established only when the satellite is visible to the ground node. Additionally, each satellite can accommodate up to $\lambda$ transmission units to build ISLs due to the limited number of antennas.

In our system, we assume that the upstream and downstream share the same bandwidth capacity for each link. The corresponding ingress relay server receives the upstream flow from a user $u$ and subsequently sends the total of flows from other attendees back to $u$ as the downstream flow. We represent the bandwidth capacity of the link $(m,n)$ as $Cap(m,n)$. Furthermore, we denote $\mathbb{B}_u^{up}$ and $\mathbb{B}_u^{down}$ as the upstream and downstream bandwidth requirements of user $u$.

\subsection{User-perceived Communication Latency Model}
Denote $x_{mn}$ as a binary variable. $x_{mn}=1$ if and only if both node $m$ and node $n$ are selected as relays, establishing a link for data transmission between the two nodes. Also, denote $y_{ij}^{mn}$ as a binary variable. $y_{ij}^{mn}=1$ if and only if data is transmitted from user $i$ to user $j$ through the link $(m,n)$.

We make the assumption that the latency from node $m$ to node $n$, represented as $L_{mn}$, is equal in the opposite direction as well. Thus, for session $e$ at slot $t$, the latency from user $u$ to its ingress relay $R_t^{e,u}$ is: $\Sigma_{m,n\in{\mathbb S\cup\mathbb U^e}}L_{mn}\cdot x_{mn}\cdot y_{u, R_t^{e,u}}^{mn}$. The end-to-end latency between two users $i$ and $j$ can be expressed as the sum of the latencies from the two users to their corresponding relays and between the relays, which can be written as: $\Sigma_{m,n\in{\mathbb S\cup\mathbb U^e}}L_{mn}\cdot x_{mn}\cdot(y_{i, R_t^{e,i}}^{mn}+y_{R_t^{e,j} ,j}^{mn}+y_{R_t^{e,i},R_t^{e,j}}^{mn})$.

Denote $P^e$ as the number of users in session $e$. Considering the end-to-end latency between every two users, the average one-way latency for each session is calculated as:

\begin{equation}
    \label{average function}
        t_{ave}=\sum\limits_{i,j\in \mathbb U^e}\frac{\Sigma_{m,n\in\{\mathbb S\cup\mathbb U^e\}}L_{mn}\cdot x_{mn}\cdot(y_{i, R_t^{e,i}}^{mn}+y_{R_t^{e,j} ,j}^{mn}+y_{R_t^{e,i},R_t^{e,j}}^{mn})}{2\cdot \tbinom{P^e}{2}}
\end{equation}
where $R_t^{e,i} ,R_t^{e,j}$ are the corresponding ingress servers of users $i,j$.

\subsection{Problem Formulation}\label{equations}

The goal is to obtain a balance between synchronization and average latency should be reached with a weight parameter $\alpha$:

    \begin{align}
        \label{goal function}\min & \sum\limits_{e\in\mathbb E}\quad (t_{ave}+\frac{\alpha}{P^e}\sum_{i=1}^{P^e}|t_{ave}-t_i|)\\             
        \text{s.t.} \label{alpha cons}&\sum\limits_{n\in\mathbb S}x_{mn}\leq\lambda, \forall m\in\mathbb S\\
        \label{vis cons}&y_{ij}^{mn}, y_{ij}^{nm}\leq x_{mn}\leq Vis(m, n), \forall m, n, i, j\in\{\mathbb S\cup\mathbb U^e\}\\
        \label{beta cons1}&\sum\limits_w y_{u,R_t^e}^{\gamma w}-\sum\limits_v y_{u,R_t^e}^{v\gamma}=\begin{cases}
        1 & \gamma=u\\ -1 & \gamma=R_t^e\\ 0 & otherwise\end{cases}\\
        \label{beta cons2}&\sum\limits_w y_{R_t^e, u}^{\gamma w}-\sum\limits_v y_{R_t^e, u}^{v\gamma}=\begin{cases}
        1 & \gamma=R_t^e\\ -1 & \gamma=u\\ 0 & otherwise\end{cases}\\
         \label{cap cons}&\sum\limits_e\sum\limits_{u, R_t^e}(\mathbb B_u^{up}\cdot y_{u,R_t^e}^{mn}+\mathbb B_u^{down}\cdot y_{R_t^e, u}^{mn})\leq Cap(m, n)   
    \end{align}

Constraint Eq. \ref{alpha cons} demonstrates that the number of ISLs for each satellite must be less than the number of transmission units. Constraint Eq. \ref{vis cons} indicates that a link can be established only if the nodes are visible
to each other, and data can be transmitted through this
link only if it is activated. Constraints Eq. \ref{beta cons1}, \ref{beta cons2} show that only one upstream flow and one downstream flow could exist between user $u$ and the control unit $R_t^e$, where $w,v,\gamma\in\{\mathbb S\cup\mathbb U^e\}$, $R_t^e\in\mathbb S$, $u\in\mathbb U^e$. Constraint Eq. \ref{cap cons} guarantees that each link could satisfy its bandwidth capacity.

\section{Algorithm Design}\label{method}

\subsection{Algorithm Overview}
The key idea involves time division into dynamic slots and managing the selection of ingress servers along with data transmission paths between them. During runtime, the virtual interaction session follows these steps: First, the system employs the ingress relay selection algorithm (Algorithm \ref{alg1}) to determine the appropriate ingress server from the available satellites. Next, the selected ingress servers receive virtual interaction requests from users within their respective regions. Subsequently, the ingress servers transmit data to each other, utilizing the flow allocation algorithm (Algorithm \ref{alg3}) to choose optimal paths between them and allocate the flow.

\subsection{Ingress Relay Selection Algorithm}
\begin{algorithm}[!tbp]
	\renewcommand{\algorithmicrequire}{\textbf{Input:}}
	\renewcommand{\algorithmicensure}{\textbf{Output:}}
	\caption{Ingress Relay Selection Algorithm}
	\label{alg1}
	\begin{algorithmic}[1]
		\STATE Initialization: $Node\leftarrow \mathbb S, R_0^{e}\leftarrow\varnothing, \forall e\in \mathbb{E}, u\in\mathbb{U}^e$
            \FOR{t=1,2,3,...,T}
                \STATE /* \textit{Establish the links.} */
                \STATE\texttt{Update(\textrm{$Vis$})} in slot $t$
                \FOR{\textbf{each} new attendee $u\in\mathbb{U}^e$}
                \STATE $\mathbb{U}^e.$\texttt{addUser(\textrm{$u$})}
                \ENDFOR
                \STATE $G\leftarrow$\texttt{UpdateGraph(\textrm{$Node\cup\mathbb{U}, Vis$})}
                \FOR{\textbf{each} $e\in\mathbb{E}$}
                \STATE /* \textit{Devide users into different regions.} */
                \STATE $region\_set\leftarrow$\texttt{Region\_Divider(\textrm{$\mathbb{U}^e, G$})}
                \FOR{\textbf{each} $re\in region\_set$}
                \STATE /* \textit{Obtain top-k nodes close to the geo-central.} */
                \STATE $cu\_set\leftarrow$\texttt{Top\_k(\textrm{$\mathbb{U}^{e,re}, G$})}
                \STATE $R_t^{e,re}\leftarrow\arg\min_{cu\in cu\_set}\{\mathbb{L}(cu, \mathbb{U}^{e,re})\}$
                \STATE /* \textit{Switch the control unit in the current slot.} */
                \IF{$\|R_t^{e,re}-R_{t-1}^{e,re}\|\geq\Delta$ \textbf{or} $\exists$ \textit{new attendee}}
                \FOR{\textbf{each} $u\in\mathbb{U}^{e,re}$}
                \STATE /* \textit{Allocate flows.} */
                \STATE \texttt{Flow(\textrm{$u, R_t^{e,re}$})}$\leftarrow$\texttt{allocate(\textrm{$G, u, $\\ $R_t^{e,re}, \mathbb{B}_u^{up}$})} 
                \STATE \texttt{Flow(\textrm{$R_t^{e,re}, u$})}$\leftarrow$\texttt{allocate(\textrm{$G, R_t^{e,re},$\\ $u, \mathbb{B}_u^{down}$})}
                \ENDFOR
                \FOR{\textbf{each other} $R_t^{e,re_2}\in R_t^e$}
                \STATE \texttt{Flow(\textrm{$R_t^{e,re},R_t^{e,re_2}$})}
                $\leftarrow$\texttt{allocate(\textrm{$G, R_t^{e,re},R_t^{e,re_2},\mathbb{B}_{re}^{re_2}$})} 
                \STATE \texttt{Flow(\textrm{$R_t^{e,re_2},R_t^{e,re}$})}
                $\leftarrow$\texttt{allocate(\textrm{$G, R_t^{e,re_2},R_t^{e,re},\mathbb{B}_{re_2}^{re}$})} 
                \ENDFOR
                \ELSE \STATE $R_t^{e,re}\leftarrow R_{t-1}^{e,re}$
                \ENDIF
                \ENDFOR
                \ENDFOR
            \ENDFOR
    \end{algorithmic}  
\end{algorithm}

Algorithm \ref{alg1} outlines the ingress relay selection algorithm in detail. For each session $e$ at time $t$, the ingress relay set is denoted as $R_t^{e}$. At every time slot, the $visibility$ and the graph $G$ is updated. All satellites and users are represented as nodes in $G$, and an edge exists between two nodes if and only if they are visible to each other. The users are then divided into different regions using the \texttt{Region\_Divider} function, ensuring that each region contains at most $N_{max}$ users and the distance between any two users is at most $d_{max}$. Each region $re$ corresponds to a specific ingress relay $R_t^{e,re}$ at time $t$.

To identify the ideal ingress relay for each region, the algorithm selects $k$ potential relays using the \texttt{Top\_k} function, filtering $k$ nodes among satellites that are closest to the center of all users' locations in that region. Subsequently, the algorithm chooses the best potential ingress server for the current slot based on its low latency synchronization ability. The parameter $\mathbb{L}(cu, \mathbb{U}^{e,re})$ represents the weighted sum of average latency and variance for session $e$ in the region $re$ according to Eq. \ref{goal function}.

In cases where the distance between two ideal ingress relay servers of the same region exceeds the threshold $\Delta$, or when a new attendee joins the session, a handover process is triggered. The \texttt{Flow(\textrm{$u, R_t^{e,re}$})} function updates the flow situation in the graph, including variables like $y_{u,R_t^{e,re}}^{mn}$. Additionally, the \texttt{allocate} function determines the best path between the user and the relay server, which will be further discussed in Algorithm \ref{alg3}. Similar procedures are followed to generate data transmission paths and flow between all ingress relays. The parameter $\mathbb{B}_{re}^{re_2}$ denotes the total bandwidth from ingress relay $R^{e,re}$ to ingress relay $R^{e, re_2}$.

\subsection{Flow Allocation Algorithm}
Algorithm \ref{alg3} presents the flow allocation algorithm in detail. The primary objective is to identify a path that offers the highest synchronization while maintaining relatively low latency. It should be noted that the number of inter-satellite links (ISLs) cannot exceed the number of transmission units on the satellite, making paths with the maximum activated links a priority. In this algorithm, the latency is determined solely by the number of hops. Paths with more hops between two nodes result in higher latency. It is also assumed that there may be several potential paths with the same number of hops between two nodes.

The algorithm initially employs Dijkstra's Algorithm to search for alternative paths with the same minimum number of hops between node $i$ and node $j$ in graph $G$. If any of these paths fail to meet the bandwidth capacity or ISL number requirements specified in \ref{equations}, they are skipped and removed from consideration. Subsequently, the algorithm counts the number of activated links for each path and selects the path that covers the maximum number of activated links as the optimal path denoted by $best$.

Finally, the algorithm activates all inactivated links in the chosen $best$ path, updates the remaining bandwidth, and allocates the flows between nodes $i$ and $j$. The result is an efficient allocation of flows that meet the desired criteria.

\begin{algorithm}[!tbp]
	\renewcommand{\algorithmicrequire}{\textbf{Input:}}
	\renewcommand{\algorithmicensure}{\textbf{Output:}}
	
	\caption{Flow Allocation Algorithm}
	\label{alg3}
	\begin{algorithmic}[1]
            \STATE \textbf{Function} \texttt{allocate(\textrm{$G, i, j, \mathbb{B}$})}
            \STATE\texttt{Flow(\textrm{i, j})}$\leftarrow\varnothing$, $all\_path \leftarrow$\texttt{DijkShortest(\textrm{$G, i, j$})}
            \FOR{\textbf{each} $path\in all\_path$}
            \STATE $path$\texttt{.ActivatedNum} $\leftarrow 0$
            \FOR{\textbf{each} $(m, n)\in path$}
            \IF{$\mathbb B_{mn}^{remain}\leq\mathbb B$ or $\sum_{k\in\mathbb S}x_{mk}\geq\lambda$ or $\sum_{k\in\mathbb S}x_{nk}\geq\lambda$}
            \STATE \textbf{break, remove $path$ from $all\_path$}
            \ENDIF
            \IF{$x_{mn}=1$}
            \STATE $path$\texttt{.ActivatedNum}++
            \ENDIF
            \ENDFOR
            \ENDFOR
            \STATE $best\leftarrow\arg\max_{p\in all\_path}(p\texttt{.ActivatedNum})$
            \FOR{\textbf{each} $(m, n)\in best$}
            \STATE $x_{mn}=1$, $y_{ij}^{mn}=1$, $\mathbb B_{mn}^{remain}=\mathbb B_{mn}^{remain}-\mathbb B$
            \STATE \texttt{Flow(\textrm{$m, n$}).Update(\textrm{$y_{ij}^{mn}$})}
            \STATE $G$\texttt{.Update(\textrm{$x_{mn},\mathbb B_{mn}^{remain}$})}
            \ENDFOR
            \STATE \textbf{Return} \texttt{Flow(\textrm{$i, j$})}
	\end{algorithmic}  
\end{algorithm}

\section{Evaluations}\label{results}
\subsection{Experiment Setup}
\subsubsection{Setup for LEO satellite constellations}

We use the simulator \cite{starperf simulator} to run STK \cite{STK} and obtain the trajectories of the Starlink (Phase 1) constellation. This constellation comprises 24 orbits and 1584 LEO satellites positioned at an altitude of approximately 550km.

In accordance with the system model defined in Section \ref{system model}, we set $\lambda=4$, $k=5$, and $\Delta=1000km$ for inter-satellite links (ISLs). The bandwidth capacities of ISLs and user-satellite links (USLs) are set to 10Gbps and 5Mbps, respectively. Each user's upstream bandwidth in a session is randomly initialized within the range of 2Mbps to 4Mbps. As for the division of regions, we set $N_{max}$ to 50 and $d_{max}$ to 1000km.

\subsubsection{User traffic generation}\label{dataset_RTC}
To better simulate real-world scenarios, we generate 5000 users on the Earth's lands, taking into account the global population distribution \cite{population dist}. These users are randomly assigned to either an ongoing session or a new session is created for them. The entry time of each user into the session is also randomly generated.

\subsection{Comparison Approaches}\label{compare}
As for comparison, we consider two other benchmarks, SpaceRTC and VIA.

\begin{figure}[!tbp]
    \centering
    \subfigure[CDF for latency of different schemes.]
    {
        \includegraphics[width=3.0in]{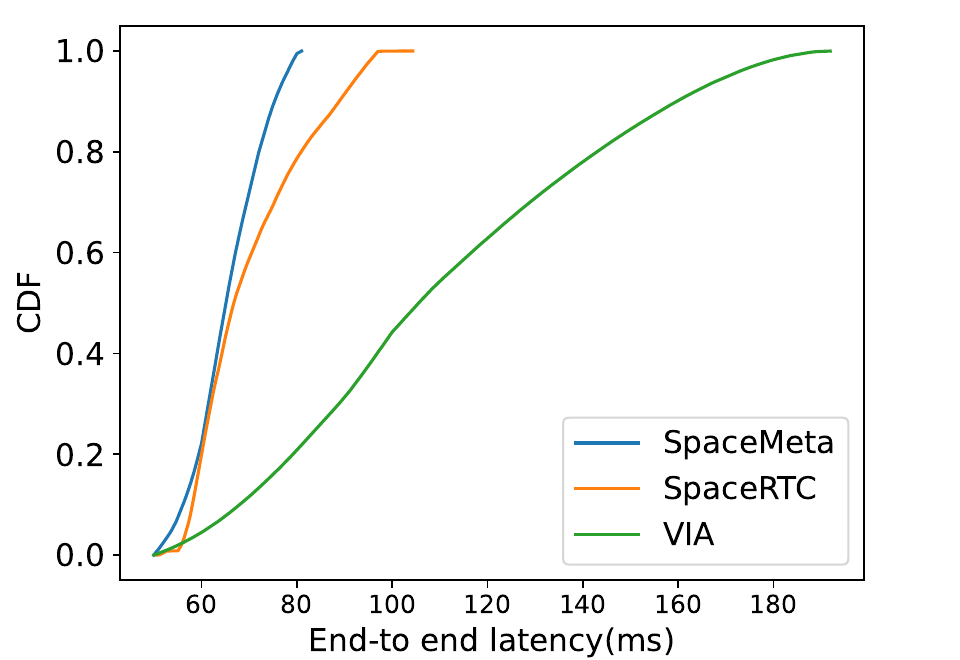}
        \label{CDF_new}
    }
    \subfigure[Boxplot for latency of different schemes.]
    {
        \includegraphics[width=3.0in]{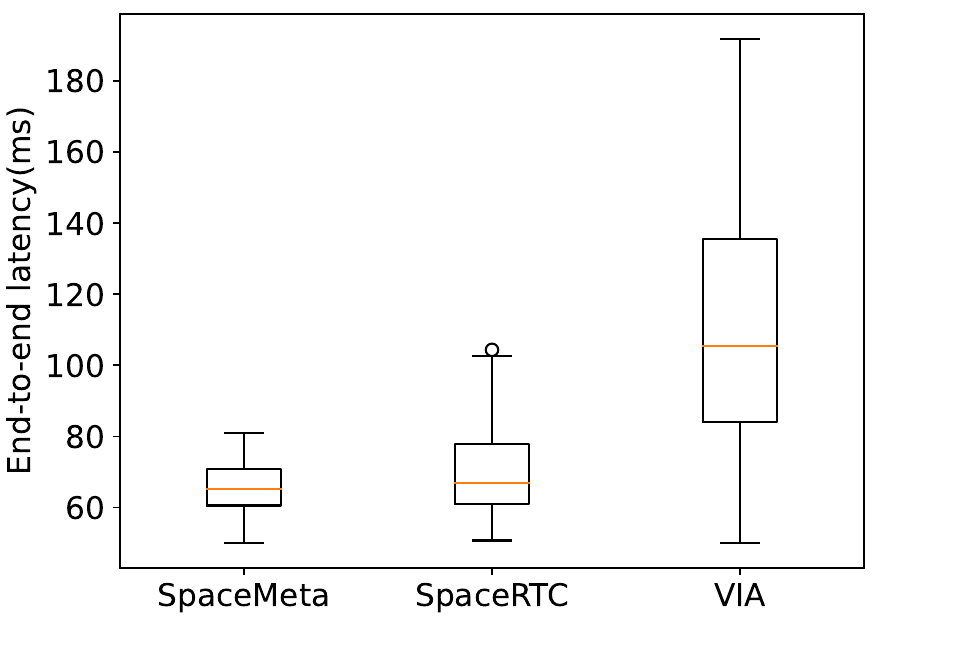}
        \label{box_new}
    }
    \caption{Performance comparison of SpaceMeta with other benchmarks.}
    \label{results fig}
\end{figure}

\begin{itemize}
    \item \textbf{SpaceRTC}\cite{SpaceRTC} is a cloud-satellite coorperated solution for low-latency problems. One control unit among satellites and cloud bases is selected for each session, and flow is generated through the path with the lowest latency and most activated links.
    \item \textbf{VIA}\cite{VIA} is a purely cloud-based scheme based on prior state-of-the-art cloud-relay selection study. It uses history performance to predict performance and gains the most promising top-$k$ control unit options. It then selects the best one and stores the results back in the session history.   
\end{itemize}

\subsection{Performance Comparison of Latency and Synchronization}

We set the weight parameter $\alpha$ in Eq. \ref{goal function} to 5 and then examine the end-to-end latency between every two users, comparing the results with SpaceRTC and VIA.

The CDF of latency is shown in Figure \ref{CDF_new}. On average, SpaceMeta reduces latency by 6.72\% compared to SpaceRTC. The results are quite close because both methods consider paths on satellites. The slightly lower latency of SpaceMeta is attributed to its multi-server selection scheme, which shortens flow paths between some distant users. Additionally, SpaceMeta reduces latency by 40.67\% on average compared to VIA. This improvement can be mainly attributed to the larger number of LEO constellation sites and higher space utilization rate in SpaceMeta, making it more conducive to optimal route planning.

The boxplot of latency is depicted in Figure \ref{box_new}. SpaceMeta reduces the Interquartile Range of latency by 39.50\% and 80.28\% compared with SpaceRTC and VIA, indicating that SpaceMeta achieves higher synchronization. The result validates the effectiveness of SpaceMeta's region-division method, as it highly synchronizes users in the same region.

\subsection{Impact of Synchronization-latency Weight Parameter}
\begin{figure}[!tbp]
    \centering
    \includegraphics[width=3.0in]{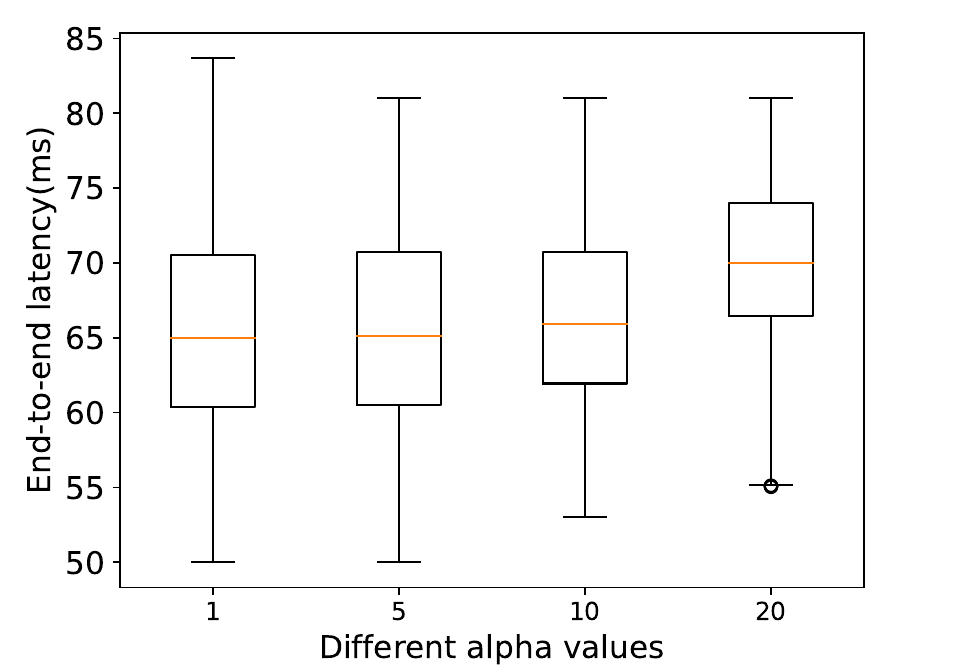}
    \caption{Latency performance with different values of $\alpha$.}
    \label{different_new}
\end{figure}

To further assess the performance of SpaceMeta under different conditions, we vary the weight parameter $\alpha$ to different values. Specifically, we compare the latency distributions for $\alpha=1, 5, 10, 20$. As illustrated in Figure \ref{different_new}, a larger value of $\alpha$ leads to higher synchronization but higher average latency. The reason behind this phenomenon is that Algorithm \ref{alg1} tends to prefer ingress relays with similar path lengths rather than shorter ones. Thus, a trade-off is needed between achieving high synchronization and maintaining low latency. It is also evident that setting $\alpha=5$ strikes a good balance between high synchronization and low latency.

\section{Conclusion}\label{conclusion}
The emergence of LEO satellite constellations presents a promising solution to support multi-user virtual interaction, enabling low latency and global coverage. In this paper, we study the server selection problem for global-scale multi-user virtual interaction applications over LEO satellite constellations. We present SpaceMeta to jointly optimize latency and synchronization, represented as latency discrepancy, among users. 
A greedy algorithm is designed to determine ingress servers and data transmission paths for each session. Experiments on real-world scale LEO constellations demonstrate that SpaceMeta can reduce the latency by up to 40.67\% on average and reduce the interquartile range of latency by up to 80.28\% when compared to benchmarks. Our study sheds light on the promising direction of supporting global-scale massive multi-user virtual interaction, also known as metaverse, via LEO satellite constellations. 

\ifCLASSOPTIONcaptionsoff
  \newpage
\fi

\end{document}